\RequirePackage{ifpdf}
\ifpdf 
\documentclass[pdftex]{sigma}
\else
\documentclass{sigma}
\fi

\def\openone{\leavevmode\hbox{\small1\kern-3.3pt\normalsize1}}

\def\bbbe{{\Bbb E}}
\def\bbbr{{\Bbb R}}
\def\bbbd{{\Bbb D}}
\def\bbbc{{\Bbb C}}
\def\bbbz{{\Bbb Z}}
\def\wedgecomma{\mathop{\wedge}\limits_{'}}
\def\otimescomma{\mathop{\otimes}\limits_{'}}

\def\rank{\mbox{rank\,}}

\def\conf{\mbox{Conf\,}}

\def\Aut{\mbox{Aut\,}}
\def\re{\mbox{Re\,}}

\font\dynkfont=cmsy10 scaled\magstep4    \skewchar\dynkfont='60
\def\dynk{\textfont2=\dynkfont}
\def\hr#1,#2;{\dimen0=.4pt\advance\dimen0by-#2pt
              \vrule width#1pt height#2pt depth\dimen0}
\def\vr#1,#2;{\vrule height#1pt depth#2pt}
\def\blb#1#2#3#4#5
            {\hbox{\ifnum#2=0\hskip11.5pt
           \else\ifnum#2=1\hr13,5.4;\hskip-1.5pt
           \else\ifnum#2=2\hr13.5,7.6;\hskip-13.5pt
                        \hr13.5,3.2;\hskip-2pt
                   \else\ifnum#2=3\hr13.7,8.4;\hskip-13.7pt
                                  \hr13,5.4;\hskip-13pt
                                  \hr13.7,2.4;\hskip-2.2pt
                   \else\ifnum#2=7\hr13.7,8.4;\hskip-13.7pt
                                  \hr13.2,6.4;\hskip-13.2pt
                                  \hr13.2,4.4;\hskip-13.2pt
                                  \hr13.7,2.4;\hskip-2.2pt
                   \else\ifnum#2=8\hr13.7,8.4;\hskip-13.7pt
                                  \hr13.2,6.4;\hskip-13.2pt
                                  \hr13.2,4.4;\hskip-13.2pt
                 \hr13.7,2.4;\hskip-8.2pt$\rangle$\hskip-2.0pt
                   \else\ifnum#2=4\hr13.5,7.6;\hskip-13.5pt
                \hr13.5,3.2;\hskip-8pt$\rangle$\hskip-1.8pt
                   \else\ifnum#2=5\hr30,5.4;\hskip-1.5pt
                   \else\ifnum#2=6\hr13.7,8.4;\hskip-13.7pt
                                  \hr13,5.4;\hskip-13pt
               \hr13.7,2.4;\hskip-8.2pt$\rangle$\hskip-2.0pt
                      \fi\fi\fi\fi\fi\fi\fi\fi\fi
                   $#1$
                   \ifnum#4=0
                   \else\ifnum#4=1\hskip-9.2pt\vr22,-9;\hskip8.8pt
                   \else\ifnum#4=2\hskip-10.9pt\vr22,-8.75;\hskip3pt
                                  \vr22,-8.75;\hskip7.1pt
                   \else\ifnum#4=3\hskip-12.6pt\vr22,-8.5;\hskip3pt
                                  \vr22,-9;\hskip3pt
                                  \vr22,-8.5;\hskip5.4pt
                   \else\ifnum#4=5\hskip-9.2pt\vr39,-9;\hskip8.8pt
                                  \fi\fi\fi\fi\fi
                   \ifnum#5=0
                   \else\ifnum#5=1\hskip-9.2pt\vr1,12;\hskip8.8pt
                   \else\ifnum#5=2\hskip-10.9pt\vr1.25,12;\hskip3pt
                                  \vr1.25,12;\hskip7.1pt
                   \else\ifnum#5=3\hskip-12.6pt\vr1.5,12;\hskip3pt
                                  \vr1,12;\hskip3pt
                                  \vr1.5,12;\hskip5.4pt
                   \else\ifnum#5=5\hskip-9.2pt\vr1,29;\hskip8.8pt
                                   \fi\fi\fi\fi\fi

                   \ifnum#3=0\hskip8pt
                   \else\ifnum#3=1\hskip-5pt\hr13,5.4;
                   \else\ifnum#3=2\hskip-5.5pt\hr13.5,7.6;
                                  \hskip-13.5pt\hr13.5,3.2;
                   \else\ifnum#3=3\hskip-5.7pt\hr13.7,8.4;
                                  \hskip-13pt\hr13,5.4;
                                  \hskip-13.7pt\hr13.7,2.4;
                  \else\ifnum#3=7\hskip-5.7pt\hr13.7,8.4;
                                 \hskip-13.2pt\hr13.2,6.4;
                                 \hskip-13.2pt\hr13.2,4.4;
                                  \hskip-13.7pt\hr13.7,2.4;
                  \else\ifnum#3=8\hskip-5.7pt\hr13.7,8.4;
                                 \hskip-13.2pt\hr13.2,6.4;
                                 \hskip-13.2pt\hr13.2,4.4;
                    \hskip-13.9pt$\langle$\hskip-8pt\hr13.7,2.4;
                   \else\ifnum#3=4\hskip-5.5pt\hr13.5,7.6;
                                  \hskip-13.5pt$\langle$\hskip-8.2pt
                                  \hr13.5,3.2;
                  \else\ifnum#3=5\hskip-5pt\hr30,5.4;
                   \else\ifnum#3=6\hskip-5.7pt\hr13.7,8.4;
                                  \hskip-13pt\hr13,5.4;
                       \hskip-13.9pt$\langle$\hskip-8.0pt\hr13.7,2.4;
                                 \fi\fi\fi\fi\fi\fi\fi\fi\fi
                   }}
\def\blob#1#2#3#4#5#6#7{\hbox
{$\displaystyle\mathop{\blb#1#2#3#4#5 }_{#6}\sp{#7}$}}
\def\up#1#2{\dimen1=33pt\multiply\dimen1by#1
                  \hbox{\raise\dimen1\rlap{#2}}}
\def\uph#1#2{\dimen1=17.5pt\multiply\dimen1by#1
                  \hbox{\raise\dimen1\rlap{#2}}}
\def\dn#1#2{\dimen1=33pt\multiply\dimen1by#1
                   \hbox{\lower\dimen1\rlap{#2}}}
\def\dnh#1#2{\dimen1=17.5pt\multiply\dimen1by#1
                    \hbox{\lower\dimen1\rlap{#2}}}

\def\rlbl#1{\kern-8pt\raise3pt\hbox{$\scriptstyle #1$}}
\def\llbl#1{\raise3pt\llap{\hbox{$\scriptstyle #1$\kern-8pt}}}
\def\elbl#1{\kern3pt\lower4.5pt\hbox{$\scriptstyle #1$}}
\def\lelbl#1{\rlap{\hbox{\kern-9pt\raise2.5pt\hbox{{$\scriptstyle #1$}}}}}

\def\wht#1#2#3#4{\blob\circ#1#2#3#4{}{}}

\def\whtd#1#2#3#4#5{\blob\circ#1#2#3#4{#5}{}}

\def\whtu#1#2#3#4#5{\blob\circ#1#2#3#4{}{#5}}
\def\blku#1#2#3#4#5{\blob\bullet#1#2#3#4{}{#5}}
\def\whtr#1#2#3#4#5{\blob\circ#1#2#3#4{}{}\rlbl{#5}}

\def\whtl#1#2#3#4#5{\llbl{#5}\blob\circ#1#2#3#4{}{}}

\def\rwng{\hbox{$\vbox{\offinterlineskip{
  \hbox{\phantom{}\kern6pt{$\circ$}}\kern-2.5pt\hbox{$\Biggr/$}\kern-0.5pt
  \hbox{\phantom{}\kern-5pt$\circ$}\kern-3.0pt\hbox{$\Biggr\backslash$}
  \kern-1.5pt\hbox{\phantom{}\kern6pt{$\circ$}} }}$}}

\def\lwng{\hbox{$\vbox{\offinterlineskip{ \hbox{$\circ$}
  \kern-3.0pt\hbox{\phantom{}\kern6.0pt{$\Biggr\backslash$}}
  \kern-0.5pt\hbox{\phantom{}\kern11pt{$\circ$}}\kern-3.5pt
  \hbox{\phantom{}\kern5.0pt {$\Biggr/$}}\kern-1.0pt\hbox{$\circ$} }}$}}

\def\drwng#1#2#3{\hbox{$\vcenter{ \offinterlineskip{
  \hbox{\phantom{}\kern6pt{$\circ^{\elbl{#3}}$}}
  \kern-2.5pt\hbox{$\Biggr/$}\kern-0.5pt
  \hbox{\phantom{}\kern-5pt$\circ^{ \elbl{#1}}$}
  \kern-3.0pt\hbox{$\Biggr\backslash$}
  \kern-1.5pt\hbox{\phantom{}\kern6pt{$\circ^{\elbl{#2}}$}}  } }$}}

\def\dlwng#1#2#3{\hbox{$\vcenter{\offinterlineskip{ \hbox{$\lelbl{#1}\circ$}
  \kern-3.0pt\hbox{\phantom{}\kern6.0pt{$\Biggr\backslash$}}
  \kern-0.5pt\hbox{\phantom{}\kern11pt{$\lelbl{#2}\circ$}}\kern-3.5pt
  \hbox{\phantom{}\kern5.0pt {$\Biggr/$}}\kern-1.0pt%
   \hbox{$\lelbl{#3}\circ$}}}$}
}

\def\rde#1#2#3{\raisebox{.5pt}{\hbox{\phantom{}\kern-4pt\hbox{$\vcenter
{\offinterlineskip\hbox{
               \raise 4.5pt\hbox{\vrule height0.4pt width13pt depth0pt}
                \kern-1pt\vbox{ \hbox{\drwng{#1}{#2}{#3}}} }}$  }} }}

\def\lde#1#2#3{\raisebox{.5pt}{\hbox{$\vcenter{\offinterlineskip  \hbox{
         \dlwng{#1}{#2}{#3}\kern-5.2pt\lower0.4pt\hbox{$\vcenter{\hrule
 width13pt}$}
               \kern-8pt\phantom{}   }}  $}}}

\def\ldet#1#2#3{\hbox{$\vcenter{\offinterlineskip  \hbox{
               \dlwng{#1}{#2}{#3}\kern14pt\lower0.4pt\hbox{$\vcenter{
          \hskip-20pt\hr13.5,5.7;\hskip-13.5pt\hr13.5,1.3;}$}
               \kern-25.8pt\phantom{}   }}  $}}

\def\rwngb{\hbox{$\vbox{\offinterlineskip{
\hbox{\phantom{}\kern6pt{$\bullet$}}\kern-2.5pt\hbox{$\Biggr/$}\kern-0.5pt
  \hbox{\phantom{}\kern-5pt$\bullet$}\kern-3.0pt\hbox{$\Biggr\backslash$}
  \kern-1.5pt\hbox{\phantom{}\kern6pt{$\bullet$}} }}$}}

\def\lwngb{\hbox{$\vbox{\offinterlineskip{ \hbox{$\bullet$}
  \kern-3.0pt\hbox{\phantom{}\kern6.0pt{$\Biggr\backslash$}}
  \kern-0.5pt\hbox{\phantom{}\kern11pt{$\bullet$}}\kern-3.5pt
  \hbox{\phantom{}\kern5.0pt {$\Biggr/$}}\kern-1.0pt\hbox{$\bullet$} }}$}}

\def\dbrwng#1#2#3{\hbox{$\vcenter{ \offinterlineskip{
  \hbox{\phantom{}\kern6pt{$\bullet^{\elbl{#3}}$}}
  \kern-2.5pt\hbox{$\Biggr/$}\kern-0.5pt
  \hbox{\phantom{}\kern-5pt$\bullet^{ \elbl{#1}}$}
  \kern-3.0pt\hbox{$\Biggr\backslash$}
  \kern-1.5pt\hbox{\phantom{}\kern6pt{$\bullet^{\elbl{#2}}$}}  } }$}}

\def\dblwng#1#2#3{\hbox{$\vcenter{\offinterlineskip{
   \hbox{$\lelbl{#1}\bullet$}
  \kern-3.0pt\hbox{\phantom{}\kern6.0pt{$\Biggr\backslash$}}
  \kern-0.5pt\hbox{\phantom{}\kern11pt{$\lelbl{#2}\bullet$}}\kern-3.5pt
  \hbox{\phantom{}\kern5.0pt
 {$\Biggr/$}}\kern-1.0pt\hbox{$\lelbl{#3}\bullet$}}}$} }

\def\rbde#1#2#3{\hbox{\phantom{}\kern-4pt\hbox{$\vcenter{\offinterlineskip
 \hbox{
               \raise 4.5pt\hbox{\vrule height0.4pt width13pt depth0pt}
                \kern-1pt\vbox{ \hbox{\dbrwng{#1}{#2}{#3}}} }}$  }}  }

\def\lbde#1#2#3{\hbox{$\vcenter{\offinterlineskip  \hbox{
        \dblwng{#1}{#2}{#3}\kern-4.2pt\lower0.4pt\hbox{$\vcenter{\hrule
 width13pt}$}
               \kern-8pt\phantom{}   }}  $}}

\def\eddgiu#1.#2.#3.{\dynk \whtu0100{#1}\whtu1300{#2}\whtu6000{#3}}
\def\eddgid#1.#2.#3.{\dynk \whtd0100{#1}\whtd1300{#2}\whtd6000{#3}}
\def\eddgiid#1.#2.#3.{\dynk  \whtd0300{#1}\whtd6100{#2}\whtd1000{#3}}

\def\eddfiu#1.#2.#3.#4.#5.{\dynk
 \whtu0100{#1}\whtu1100{#2}\whtu1200{#3}\whtu4100{#4}\whtu1000{#5}}

\def\eddfid#1.#2.#3.#4.#5.{\dynk
 \whtd0100{#1}\whtd1100{#2}\whtd1200{#3}\whtd4100{#4}\whtd1000{#5}}

\def\eddfiiu#1.#2.#3.#4.#5.{\dynk
 \whtu0100{#1}\whtu1200{#2}\whtu4100{#3}\whtu1100{#4}\whtu1000{#5}}
\def\eddfiid#1.#2.#3.#4.#5.{\dynk
 \whtu0100{#1}\whtd1200{#2}\whtd4100{#3}\whtu1100{#4}\whtd1000{#5}}

\def\ddanu#1.#2.#3.#4.#5.{\dynk \whtu0100{#1}\whtu1100{#2}\whtu1100{#3}%
                          \cdots\whtu1100{#4}\whtu1000{#5}}
\def\ddanuf#1.#2.#3.#4.{\dynk \whtd0100{#1}\whtu1100{#2}\cdots%
                           \whtu1100{#3}\whtu1000{#4}}
\def\ddanuuf#1.#2.#3.#4.{\dynk \whtu0100{#1}\whtu1100{#2}\cdots%
                           \whtu1100{#3}\whtu1000{#4}}
\def\ddanufd#1.#2.#3.#4.{\dynk \whtd0100{#1}\whtu1100{#2}\cdots%
                           \whtu1100{#3}\whtr1005{#4}}
\def\ddandf#1.#2.#3.#4.{\dynk \whtu0100{#1}\whtd1100{#2}\cdots%
                           \whtd1100{#3}\whtd1000{#4}}
\def\ddanddf#1.#2.#3.#4.{\dynk \whtd0100{#1}\whtd1100{#2}\cdots%
                           \whtd1100{#3}\whtd1000{#4}}
\def\ddandfu#1.#2.#3.#4.{\dynk \whtu0100{#1}\whtd1100{#2}\cdots%
                           \whtd1100{#3}\whtr1050{#4}}
\def\ddcnds#1.#2.#3.#4.#5.#6{\dynk \whtd0200{#1}\whtd4100{#2}%
                          \whtd1100{#3}\cdots%
                           \whtd1100{#4}\whtd1400{#5}\whtd2000{#6}}

\def\eddanu#1.#2.#3.#4.#5.{\dynk \whtu0100{#1}\whtu1100{#2}%
               \up1{\whtr0000{#3}}\cdots\whtu1100{#4}\whtu1000{#5}}
\def\eddand#1.#2.#3.#4.#5.{\dynk \whtd0100{#1}\whtd1100{#2}%
                \up1{\whtr0000{#3}}\cdots\whtd1100{#4}\whtd1000{#5}}
\def\ddand#1.#2.#3.#4.#5.{\dynk \whtd0100{#1}\whtd1100{#2}\whtd1100{#3}%
               \cdots\whtd1100{#4}\whtd1000{#5}}
\def\andfive#1.#2.#3.#4.#5.{\dynk \whtu0100{#1}\whtu1100{#2}\whtu1100{#3}%
                           \whtu1100{#4}\whtu1000{#5}}
\def\andthr#1.#2.#3.{\dynk \whtu0100{#1}\whtu1100{#2}\whtu1000{#3}}

\def\eddanid#1.#2.#3.#4.#5.{\dynk \whtd0200{#1}\whtd4100{#2}%
                           \whtd1100{#3}\cdots\whtd1200{#4}\whtd4000{#5}}
\def\eddanidr#1.#2.#3.#4.#5.{\dynk \whtd0200{#1}\whtd4100{#2}%
                           \cdots\whtd1100{#3}\whtd1200{#4}\whtd4000{#5}}

\def\eddaniid#1.#2.#3.#4.#5.#6.{\hbox{$\vcenter{\hbox
         {\dynk\hbox{$ \lde{#1}{#2}{#3}\whtd1100{#4}\cdots%
          \whtd1400{#5}\whtd2000{#6} $}} }$}}

\def\eddaiii#1.#2.#3.{\dynk\whtd0400{#1}\whtd2200{#2}\whtd4000{#3}}
\def\eddaiiif#1.#2.#3.#4.{\dynk\whtd0400{#1}\whtd2100{#2}%
                    \whtd1200{#3}\whtd4000{#4}}
\def\eddciii#1.#2.#3.{\dynk\whtd0200{#1}\whtd4400{#2}\whtd2000{#3}}

\def\eddbnd#1.#2.#3.#4.#5.#6.{\dynk \lde{#1}{#2}{#3}\whtd1100{#4}\cdots%
                           \whtd1200{#5}\whtd4000{#6}}
\def\eddbndt#1.#2.#3.#4.{\dynk \ldet{#1}{#2}{#3}\hskip-444pt\whtr4000{#4}}

\def\ncddlr#1.#2.#3.#4.#5.{\dynk \whtu0101{#1}\whtu1100{#2}\whtu1100{#3}%
                           \cdots\whtu1100{#4}\whtu1001{#5}}
\def\ncdulr#1.#2.#3.#4.#5.{\dynk \whtd0110{#1}\whtd1100{#2}\whtd1100{#3}%
                           \cdots\whtd1100{#4}\whtd1010{#5}}
\def\ncddrd#1.#2.#3.#4.#5.{\dynk \whtd0100{#1}\whtu1100{#2}\whtu1100{#3}%
                           \cdots\whtu1100{#4}\whtu1005{#5}}
\def\ncddru#1.#2.#3.#4.#5.{\dynk \whtu0100{#1}\whtd1100{#2}\whtd1100{#3}%
                           \cdots\whtd1100{#4}\whtd1050{#5}}
\def\ncddrdu#1.#2.#3.#4.#5.{\dynk \whtu0100{#1}\whtu1100{#2}\whtu1100{#3}%
                         \cdots\whtu1100{#4}\whtu1005{#5}}
\def\ncddrud#1.#2.#3.#4.#5.{\dynk \whtd0100{#1}\whtd1100{#2}\whtd1100{#3}%
                            \cdots\whtd1100{#4}\whtd1050{#5}}
\def\ncddld#1.#2.#3.#4.#5.{\dynk \whtl0105{#1}\whtu1100{#2}\whtu1100{#3}%
                          \cdots\whtu1100{#4}\whtd1000{#5}}
\def\ncddlu#1.#2.#3.#4.#5.{\dynk \whtl0150{#1}\whtd1100{#2}\whtd1100{#3}%
                           \cdots\whtd1100{#4}\whtu1000{#5}}
\def\ncanur#1.#2.#3.#4.#5.{\dynk \whtu0101{#1}\whtu1100{#2}\whtu1100{#3}%
                           \cdots\whtu1100{#4}\whtd1010{#5}}
\def\ncandr#1.#2.#3.#4.#5.{\dynk \whtd0110{#1}\whtd1100{#2}\whtd1100{#3}%
                           \cdots\whtd1100{#4}\whtu1001{#5}}

\def\eddcnd#1.#2.#3.#4.#5.{\dynk \whtd0200{#1}\whtd4100{#2}\whtd1100{#3}
       \cdots \whtd1400{#4}\whtd2000{#5}}

\def\dddnu#1.#2.#3.#4.#5.#6.{\hbox{$\vcenter{\hbox
         {\dynk\hbox{$ \whtu0100{#1}\whtu1100{#2}\cdots%
          \whtu1100{#3}\rde{#4}{#5}{#6} $}}  }$}}
\def\dddnd#1.#2.#3.#4.#5.#6.{\hbox{$\vcenter{\hbox
         {\dynk\hbox{$ \whtd0100{#1}\whtd1100{#2}\cdots%
          \whtd1100{#3}\rde{#4}{#5}{#6} $}} }$}}
\def\dddiv#1.#2.#3.#4.{\hbox{$\vcenter{\hbox
         {\dynk\hbox{$ \whtu0100{#1}\rde{#2}{#3}{#4}
              $}}  }$}}
\def\edddiv#1.#2.#3.#4.{\hbox{$\vcenter{\hbox{\dynk\hbox{$\whtl0100{#1}
\up1{\whtl0001{#2}}\dn1{\whtl0010{#4}}\wht1111\whtr1000{#3}
$}}}$}}

\def\edddnu#1.#2.#3.#4.#5.#6.#7.#8.{\hbox{$\vcenter{\hbox
         {\dynk\hbox{$ \lde{#1}{#2}{#3}\whtu1100{#4}\cdots%
          \whtu1100{#5}\rde{#6}{#7}{#8} $}}  }$}}
\def\edddnd#1.#2.#3.#4.#5.#6.#7.#8.{\hbox{$\vcenter{\hbox
         {\dynk\hbox{$ \lde{#1}{#2}{#3}\whtd1100{#4}\cdots%
          \whtd1100{#5}\rde{#6}{#7}{#8} $}} }$}}
\def\edddndf#1.#2.#3.#4.#5.#6.{\hbox{$\vcenter{\hbox
         {\dynk\hbox{$ \lde{#1}{#2}{#3}\rde{#4}{#5}{#6} $}} }$}}

\def\edddnds#1.#2.#3.#4.#5.#6.#7.#8.#9.{\hbox{$\vcenter{\hbox
{\dynk\hbox{$ \lde{#1}{#2}{#3}\whtd1100{#4}\cdot\cdot\whtd1100{#5}\cdot%
      \cdot\whtd1100{#6}\rde{#7}{#8}{#9} $}} }$}}
\def\eddanod#1.#2.#3.#4.#5.#6.{\hbox{$\vcenter{\hbox
         {\dynk\hbox{$ \whtd0200{#1}\whtd4100{#2}\cdots%
          \whtd1100{#3}\rde{#4}{#5}{#6} $}} }$}}

\def\edddniid#1.#2.#3.#4.#5.{\hbox{$\vcenter{\hbox
         {\dynk\hbox{$ \whtd0400{#1}\whtd2100{#2}\whtd1100{#3}\cdots%
          \whtd1200{#4}\whtd4000{#5} $}} }$}}

\def\edddniiu#1.#2.#3.#4.#5.{\hbox{$\vcenter{\hbox
         {\dynk\hbox{$ \blku0200{#1}\whtu2100{#2}\whtu1100{#3}\cdots%
          \whtu1200{#4}\blku2000{#5} $}} }$}}

\def\ddei#1.#2.#3.#4.#5.#6.{\hbox{$\vcenter{\hbox
       {\dynk \whtd0100{#1}\whtd1100{#3}%
       \up1{\whtr0001{#2}}\whtd1110{#4}\whtd1100{#5}\whtd1000{#6}} }$}}

\def\eddei#1.#2.#3.#4.#5.#6.#7.{\hbox{$\vcenter{\hbox
       {\dynk \whtu0100{#1}\whtu1100{#3}%
       \up1{\whtr0011{#2}}\up2{\whtr0001{#7}}\whtd1110{#4}\whtu1100{#5}%
       \whtu1000{#6}} }$}}

\def\ncdddt#1.#2.{\dynk\whtu0400{#1}\whtu2001{#2}}
\def\ncandrt#1.#2.{\dynk\whtd0110{#1}\whtu1001{#2}}
\def\ncanurt#1.#2.{\dynk\whtu0101{#1}\whtd1010{#2}}
\def\ncddet#1.#2.{\dynk\whtu0400{#1}\whtu2000{#2}}
\def\ncddut#1.#2.{\dynk\whtd0400{#1}\whtd2010{#2}}
\def\ncdduot#1.#2.{\dynk\whtd0210{#1}\whtd4000{#2}}
\def\ncddct#1.#2.{\hbox{\dynk\whtu0200{\rotatebox{45}{$\scriptstyle#1$}}%
                    \whtu4000{\rotatebox{45}{$\scriptstyle#2$}}}}
\def\ncddcot#1.#2.{\dynk\whtd0400{\rotatebox{45}{$\scriptstyle#1$}}%
                    \whtd2000{\rotatebox{45}{$\scriptstyle#2$}}}
\def\ncddcst#1.#2.{\dynk\whtu0400{\rotatebox{135}{$\scriptstyle#1$}}%
                    \whtu2000{\rotatebox{135}{$\scriptstyle#2$}}}

\def\rronit#1.{\rotatebox{315}
       {\dynk\whtr5005{\rotatebox{45}{$\scriptstyle #1$}}}}
\def\laronit#1.#2.{\rotatebox{315}{\ncddct#1.#2.}}
\def\raronit#1.#2.{\rotatebox{315}{\ncddcot#1.#2.}}
\def\rarsnit#1.#2.{\rotatebox{225}{\ncddcst#1.#2.}}

\def\ncddd#1.#2.#3.#4.#5.{\dynk\whtu0400{#1}\whtu2100{#2}\whtu1100{#3}%
           \cdots\whtu1100{#4}\whtu1001{#5}}
\def\ncdde#1.#2.#3.#4.#5.{\dynk\whtu0400{#1}\whtu2100{#2}\whtu1100{#3}%
           \cdots\whtu1100{#4}\whtu1000{#5}}
\def\ncdded#1.#2.#3.#4.#5.{\dynk\whtl0400{#1}\whtd2100{#2}\whtd1100{#3}%
           \cdots\whtd1100{#4}\whtd1000{#5}}
\def\ncddeo#1.#2.#3.#4.#5.{\dynk\whtd0100{#1}\whtd1100{#2}\whtd1100{#3}%
           \cdots\whtd1200{#4}\whtd4000{#5}}
\def\ncddeof#1.#2.#3.#4.{\dynk\whtd0100{#1}\whtd1100{#2}%
           \cdots\whtd1200{#3}\whtr4000{#4}}
\def\ncddu#1.#2.#3.#4.#5.{\dynk\whtd0400{#1}\whtd2100{#2}\whtd1100{#3}%
           \cdots\whtd1100{#4}\whtd1010{#5}}
\def\ncdduo#1.#2.#3.#4.#5.{\dynk\whtd0110{#1}\whtd1100{#2}\whtd1100{#3}%
           \cdots\whtd1200{#4}\whtd4000{#5}}
\def\ncddc#1.#2.#3.#4.#5.{\dynk\whtu0200{#1}\whtu4100{#2}\whtu1100{#3}%
           \cdots\whtu1100{#4}\whtu1000{#5}}
\def\ncdddc#1.#2.#3.#4.#5.{\dynk\whtd0200{#1}\whtd4100{#2}\whtd1100{#3}%
           \cdots\whtd1100{#4}\whtd1000{#5}}
\def\ncddcu#1.#2.#3.#4.#5.{\dynk\whtl0200{#1}\whtd4100{#2}\whtd1100{#3}%
           \cdots\whtd1100{#4}\whtr1050{#5}}
\def\ncddcd#1.#2.#3.#4.#5.{\dynk\whtu0200{#1}\whtu4100{#2}\whtu1100{#3}%
           \cdots\whtu1100{#4}\whtu1005{#5}}
\def\ncddco#1.#2.#3.#4.#5.{\dynk\whtd0100{#1}\whtd1100{#2}\cdots%
            \whtd1100{#3}\whtd1400{#4}\whtd2000{#5}}
\def\ncdfr#1.#2.#3.#4.#5.#6.{\ncddc#1.#2.#3.#4.#5.
           \hskip-42.5pt\dynk\rotatebox{315}
          {\whtu5005{\rotatebox{45}{$\scriptstyle #6$}}}}
\def\ncdfrd#1.#2.#3.#4.#5.#6.{\ncdde#1.#2.#3.#4.#5.
           \hskip-42.5pt\dynk\rotatebox{315}
          {\whtr5005{\rotatebox{45}{$\scriptstyle #6$}}}}
\def\ncdfrdc#1.#2.#3.#4.#5.#6.{\ncddc#1.#2.#3.#4.#5.
           \hskip-42.5pt\dynk\rotatebox{315}
          {\whtr5005{\rotatebox{45}{$\scriptstyle #6$}}}}
\def\ncdfrdl#1.#2.#3.#4.#5.#6.{\ncddlu#1.#2.#3.#4.#5.
           \hskip-42.5pt\dynk\rotatebox{315}
          {\whtr5005{\rotatebox{45}{$\scriptstyle #6$}}}}

\def\lronit#1.{\rotatebox{315}
       {\dynk\whtl0550{\rotatebox{45}{$\scriptstyle #1$}}}}
\def\daone#1.#2.{\dynk\whtd0400{#1}\whtd4000{#2}}

\def\ncdfl#1.#2.#3.#4.#5.#6.{\hbox{\lronit#1.\hskip-39.5pt
                   \raisebox{13pt}{$\ddand#2.#3.#4.#5.#6.$}}}
\def\ncdfal#1.#2.#3.#4.#5.{\hbox{\lronit#1.\hskip-39.5pt
                   \raisebox{13pt}{$\ddandf#2.#3.#4.#5.$}}}
\def\ncdfalu#1.#2.#3.#4.#5.{\hbox{\lronit#1.\hskip-39.5pt
                   \raisebox{13pt}{$\ddandfu#2.#3.#4.#5.$}}}

\def\ncdfar#1.#2.#3.#4.#5.{\ddanuf#1.#2.#3.#4.
  \hskip-42.5pt\dynk\rotatebox{315}{\whtr5005{\rotatebox{45}
           {$\scriptstyle #5$}}}}
\def\ncdfaur#1.#2.#3.#4.#5.{\ddanuuf#1.#2.#3.#4.
  \hskip-42.5pt\dynk\rotatebox{315}{\whtu5005{\rotatebox{45}
           {$\scriptstyle #5$}}}}

\def\datwot#1.#2.{\dynk\whtu0700{#1}\whtu8000{#2}}

\def\datwon#1.#2.#3.#4.#5.#6.{\dynk \whtd0200{#1}\whtd4100{#2}%
         \whtd1100{#3}\whtd1100{#4} \cdots\whtd1200{#5}\whtd4000{#6}}
\def\datwono#1.#2.#3.#4.#5.#6.{\dynk \whtd0400{#1}\whtd2100{#2}%
        \whtd1100{#3}\cdots \whtd1100{#4}\whtd1400{#5}\whtd2000{#6}}
\def\datwonl#1.#2.#3.#4.#5.#6.{\dynk \whtd0200{#1}\whtd4100{#2}%
         \whtd1100{#3}\cdots \whtd1100{#4} \whtd1200{#5}\whtd4000{#6}}

\def\ddeii#1.#2.#3.#4.#5.#6.#7.{\hbox{$\vcenter{\hbox
       {\dynk \whtd0100{#1}\whtd1100{#3}%
       \up1{\whtr0001{#2}}\whtd1110{#4}\whtd1100{#5}\whtd1100{#6}%
       \whtd1000{#7}} }$}}

\def\eddeii#1.#2.#3.#4.#5.#6.#7.#8.{\hbox{$\vcenter{\hbox
       {\dynk \whtu0100{#8}\whtu1100{#1}\whtu1100{#3}%
       \up1{\whtr0001{#2}}\whtd1110{#4}\whtu1100{#5}\whtu1100{#6}%
       \whtu1000{#7}} }$}}

\def\ddeiii#1.#2.#3.#4.#5.#6.#7.#8.{\hbox{$\vcenter{\hbox
       {\dynk \whtd0100{#1}\whtd1100{#3}%
       \up1{\whtr0001{#2}}\whtd1110{#4}\whtd1100{#5}\whtd1100{#6}%
       \whtd1100{#7}\whtd1000{#8}} }$}}

\def\eddeiii#1.#2.#3.#4.#5.#6.#7.#8.#9.{\hbox{$\vcenter{\hbox
       {\dynk \whtd0100{#1}\whtd1100{#3}%
       \up1{\whtr0001{#2}}\whtd1110{#4}\whtd1100{#5}\whtd1100{#6}%
       \whtd1100{#7}\whtd1100{#8}\whtd1000{#9}} }$}}

\newcommand\alp{\alpha_}
\newcommand\bet{\beta_}

\begin{document}
\allowdisplaybreaks

\renewcommand{\PaperNumber}{022}

\FirstPageHeading

\ShortArticleName{Real Hamiltonian Forms of Af\/f\/ine Toda Models
Related to Exceptional Lie Algebras}

\ArticleName{Real Hamiltonian Forms of Af\/f\/ine Toda Models\\
Related to Exceptional Lie Algebras}

\Author{Vladimir S. GERDJIKOV~$^\dag$ and Georgi G.
GRAHOVSKI~$^{\dag \ddag}$}

\AuthorNameForHeading{V.S. Gerdjikov and G.G. Grahovski}

\Address{$^\dag$~Institute for Nuclear Research and Nuclear
Energy, Bulgarian Academy of Sciences,\\
$\phantom{^\dag}$~72 Tsarigradsko Chaussee, 1784 Sof\/ia,
Bulgaria}
\EmailD{\href{mailto:gerjikov@inrne.bas.bg}{gerjikov@inrne.bas.bg},
\href{mailto:grah@inrne.bas.bg}{grah@inrne.bas.bg}}

\Address{$^\ddag$~Laboratoire de Physique Th\'eorique et
Mod\'elisation,
 Universit\'e de Cergy-Pontoise,\\
$\phantom{^\ddag}$~2  Avenue Adolphe Chauvin, F-95302
Cergy-Pontoise Cedex, France}

\ArticleDates{Received December 19, 2005, in f\/inal form February
05, 2006; Published online February 17, 2006}

\Abstract{The construction of a family of real Hamiltonian forms
(RHF) for  the special class of af\/f\/ine $1+1 $-dimensional Toda
f\/ield theories (ATFT) is reported. Thus the method, proposed in
\cite{2} for systems with f\/inite number of degrees of freedom is
generalized to inf\/inite-dimensional Hamiltonian systems.  The
construction method is illustrated on the explicit nontrivial
example of RHF of ATFT related  to the exceptional algebras ${\bf
E}_6$ and ${\bf E}_7$. The involutions of the local integrals of
motion are proved by means of the classical $R$-matrix approach.}

\Keywords{solitons; af\/f\/ine Toda f\/ield theories; Hamiltonian
systems}

\Classification{37K15; 17B70; 37K10; 17B80}

\section{Introduction}\label{sec:1}

To each simple Lie algebra ${\rank g} $ one can relate Toda f\/ield
theory (TFT) in $1+1 $ dimensions. It allows Lax representation:
$[L,M]=0$, where $L $ and $M $ are f\/irst order ordinary
dif\/ferential operators, see e.g.\
\cite{Mikh,OlPerMikh,Olive,1,SasKha,Holy}:
\begin{gather}\label{eq:2.1}
L\psi \equiv  \left(  i{d  \over dx } - iq_x(x,t) - \lambda
J_0\right) \psi (x,t,\lambda )=0, \qquad q_x(x,t) = \sum_{k=1}^{r}
q_{k,x} H_k, \\
M\psi \equiv  \left(  i{d  \over dt } -  {1\over \lambda}
I(x,t)\right) \psi (x,t,\lambda )=0.\nonumber
\end{gather}
whose potentials take values in ${\frak  g} $. Here $q(x,t) \in
{\frak h}$ is the Cartan subalgebra of ${\frak g}$,
$\vec{q}(x,t)=\sum\limits_{k=1}^{r} q_k \vec{e}_k $ is its dual $r
$-component vector, $r=\mbox{rank}\,{\frak  g} $. $H_k$ are the
Cartan generators dual to the orthonormal basis elements
$\vec{e}_k$ in the root space, and
\begin{gather*}
J_0 = \sum_{\alpha \in \pi}^{} E_{\alpha },\qquad I(x,t) =
\sum_{\alpha \in \pi}^{} e^{-(\alpha ,\vec{q}(x,t))} E_{-\alpha }.
\end{gather*}
By $\pi_{\frak g} $ we denote the set of admissible roots of
${\frak g} $, i.e.\  $\pi_{{\frak g}} = \{\alpha _0, \alpha
_1,\dots, \alpha _r\} $ where $\alpha _1,\dots, \alpha _r $ are
the simple roots of ${\frak  g}$ and $\alpha _0 $ is the minimal
root of ${\frak  g} $.  The corresponding TFT is known as the
af\/f\/ine TFT. The Dynkin graph that corresponds to the set of
admissible roots of ${\frak g}$ is called extended Dynkin diagrams
(EDD). The equations of motion are of the form:
\begin{gather*}
{\partial ^2 \vec{q}  \over \partial x \partial t } =
\sum_{j=0}^{r} \alpha_j e^{-(\alpha_j ,\vec{q}(x,t))}.
\end{gather*}
 The present paper extends the
ideas of \cite{Elba} and \cite{vg} to the ATFT related to the
exceptional simple Lie algebra ${\bf E}_6 $; for f\/inite Toda
chains see \cite{Bogo,Dam1,2,GIG}.

\section{The reduction group} \label{sec:2}

The operators $L $ and $M $ are invariant with respect to the
reduction group $\mathcal{ G}_\bbbr\simeq \bbbd_h $ where $h $ is
the Coxeter number of ${\frak  g} $. It is generated by two
elements satisfying $g_1^h = g_2^2 =(g_1g_2)^2=\openone  $ which
allow realizations both as elements in $\Aut_{{\frak  g}} $ and in
$\conf \, \bbbc $. The invariance condition has the form
\cite{Mikh}:
\begin{alignat}{3}
& C_1(U(x,t,\kappa _1(\lambda ))) = U(x,t,\lambda ), \qquad &&
C_1(V(x,t,\kappa _1(\lambda ))) = V(x,t,\lambda ),& \nonumber\\
& C_k(U^\dag(x,t,\kappa _k(\lambda ))) = U(x,t,\lambda ), \qquad &&
C_k(V^\dag(x,t,\kappa _k(\lambda ))) = V(x,t,\lambda ), \qquad
k=2,3.\label{eq:3.1}
\end{alignat}
where $U(x,t,\lambda ) = -iq_x(x,t) - \lambda J_0$ and
$V(x,t,\lambda ) = -{1\over \lambda} I(x,t) $.  Here $C_k $ are
automorphisms of f\/inite order of ${\frak g} $, i.e.\
$C_1^h=C_2^2=(C_1C_2)^2=\openone  $ while $\kappa _k(\lambda ) $
are conformal mappings of the complex $\lambda $-plane:
\begin{gather*}
\kappa _1(\lambda )=\omega \lambda , \qquad \kappa _2(\lambda
)=\lambda^* , \qquad \kappa _3(\lambda )=(\omega \lambda)^* ,
\end{gather*}
where $\omega =\exp(2\pi i/h) $. The algebraic constraints
(\ref{eq:3.1}) are automatically compatible with the evolution. A
number of nontrivial reductions of nonlinear evolution equations
can be found in~\cite{Como,3}.

\section{Spectral properties of the Lax operator}\label{sec:3}

The reduction conditions (\ref{eq:3.1}) lead to  rather special
properties of the operator $L $. Along with $L $ we will use also
the equivalent system:
\begin{gather}\label{eq:L-t}
\widetilde{L}m(x,t,\lambda ) \equiv i {dm  \over dx } - iq_x
m(x,t,\lambda ) - \lambda [J_0, m(x,t,\lambda )] =0,
\end{gather}
where $m(x,t,\lambda )=\psi (x,t,\lambda )e^{iJ_0x\lambda } $.
Here $\vec{q}_x $ is the potential which we choose to be a
Schwartz-type function taking values in the complexif\/ied Cartan
subalgebra $\mathfrak{h}_\bbbc \subset \mathfrak{g} $. This means
that the boundary conditions for $\vec{q} $ are determined up to
the constant vector $\vec{\rho }_0 $:
\begin{gather*}
\vec{\rho }_0 =\lim_{x\to\pm\infty } \vec{q}(x,t) .
\end{gather*}
The change of the variables $\vec{q}\,{}' =\vec{q}- \vec{\rho }_0$
does not af\/fect the potential and the spectral data of~$\widetilde{L} $,
 but they change the right hand side of the ATFT
equations into:
\begin{gather*}
{\partial ^2 \vec{q}\,{}'  \over \partial x \partial t } =
\sum_{j=0}^{r} s_j \alpha_j e^{-(\alpha_j ,\vec{q}\,{}'(x,t))},
\end{gather*}
where $s_j=\exp[-(\alpha_j,\vec{\rho}_0)]$ obviously satisfy the
consistency conditions:
\begin{gather*}
\prod_{j=0}^{r} s_j^{n_j} = \exp \left(-\sum_{j=0}^{r} (n_j \alpha
_j ,\vec{\rho }_0)\right) =1.
\end{gather*}
where $n_j $ are the minimal positive integers for which
$\sum\limits_{j=0}^{r} n_j\alpha _j=0 $. In the literature there
are two canonical ways of f\/ixing up the vector $\vec{\rho }_0$.
The f\/irst one is to put $s_j=1 $ for all $j $, the second is to
have $s_j=n_j $ for all $j $.

The spectral properties of the Lax operator are rather involved
due to the fact that both $J_0$ and $I(x,t)$ have complex-valued
eigenvalues. In fact all these eigenvalues are constant and are
proportional to $\omega^k$, $k=0, 1, \dots , h-1$, where
$\omega=e^{2\pi i/h}$ and $h$ is the Coxeter number of ${\frak
g}$. As a result one f\/inds that the continuous spectrum of
$L(\lambda)$ f\/ills up $2h$ rays passing through the origin:
\[
\lambda \in l_\nu \, : \, \mbox{arg}\, \lambda = {(\nu -1)\pi
\over h}.
\]
For such operators one can introduce Jost solutions only for
potentials on a compact sup\-port~\cite{BeCo,GeYa}:
\begin{gather*}
\lim_{x \to \infty}\Psi (x,t,\lambda )e^{i\lambda J_0 x+(i/\lambda
) I_0 t} =\openone, \qquad \lim_{x \to -\infty}\Phi (x,t,\lambda
)e^{i\lambda J_0 x+(i/\lambda ) I_0 t} =\openone,
\end{gather*}
where
\[
I_0=\lim_{x \to \infty} I(x,t)=\sum_{k=0}^r E_{-\alpha_k}, \qquad
\alpha_k \in \pi ({\frak g}).
\]
Then the corresponding scattering matrix $T(t,\lambda )$ is
def\/ined by
\begin{gather*}
T(t,\lambda )=(\Psi (x,t,\lambda ))^{-1}\Phi (x,t,\lambda ).
\end{gather*}
The compactness of the potential ensures that  the Jost solutions
$\Phi (x,t,\lambda )$, $\Psi (x,t,\lambda )$ and the scattering
matrix $T(t,\lambda )$ are meromorphic functions of $\lambda$.

The fact that $\Phi (x,t,\lambda )$ and $\Psi (x,t,\lambda )$ are
also Jost solutions of $M(\lambda)$ means that $T(t,\lambda)$
evolves according to:
\begin{gather}\label{eq:dT/dt}
i{dT\over dt}-{1\over \lambda}[I_0,T(t,\lambda )]=0.
\end{gather}
The limiting procedure to non-compact potentials can be done only
for the so-called fundamental analytical solutions (FAS) $m_\nu
(x,t,\lambda)$, $\lambda \in \Omega_\nu$, i.e. $(\nu - 1)\pi/h
\leq \mbox{arg}\, \lambda \leq \nu \pi/h$. Their construction is
outlined in \cite{BeCo,GeYa} for generic complex-valued $J_0$. Our
Lax pair is special, because it satisf\/ies the reduction
condition (\ref{eq:3.1}) with the Coxeter automorphism \cite{Hum}:
\begin{gather*}
\bar{C}_1 (X) \equiv C_1^{-1}XC_1 , \qquad C_1 = e^{2\pi iH_\rho
/h}, \qquad \rho = {1  \over 2 }\sum_{\alpha >0}\alpha ;
\end{gather*}
obviously $C_1^h=\openone  $ and $\bar{C}_1(J_0)=\omega^{-1}J_0 $.
The FAS $m_\nu (x,t,\lambda )$ of (\ref{eq:L-t}) are def\/ined
only in the sector~$\Omega_\ni$ and satisfy:
\begin{gather*}
 \bar{C}_1 (m_\nu (x,t,\omega \lambda )) = m_{\nu-2
}(x,t,\lambda), \qquad \lambda \in l_{\nu -2}.
\end{gather*}
Thus the inverse scattering problem for $L(\lambda)$ can be
formulated as a Riemann--Hilbert problem for $m_\nu (x,t,\lambda
)$ on the continuous spectrum $\Sigma \equiv
\bigcup_{\nu=1}^{2h}l_\nu$.

Skipping the details, we remark that equation (\ref{eq:dT/dt})
allows one to show that ATFT possess generating functionals of the
integrals of motion. This can be shown easily if the
potential~$q(x,t)$ is such that $T(t,\lambda)$ can be
diagonalized:
\begin{gather}\label{eq:T-D}
T(t,\lambda)=u_0^{-1}(t,\lambda)D(\lambda)u_0(t,\lambda).
\end{gather}
All factors in the above equation take values in the Lie group
${\cal G}$ with the Lie algebra ${\frak g}$ and $D(\lambda)$ is a
diagonal matrix. Then there exist a set of functions
$\tau_k(\lambda)$, $k=1,\dots ,r$ such that:
\begin{gather}\label{eq:D-tau}
D(\lambda)=\exp \left(\sum_{k=1}^{r}{2\tau_k(\lambda)\over
(\alpha_k, \alpha_k)}H_{\alpha_k} \right),
\end{gather}
where $H_{\alpha_k}$ are the Cartan generators of ${\frak g}$
corresponding to the simple root $\alpha_k$. Inserting
(\ref{eq:T-D}) and (\ref{eq:D-tau}) into (\ref{eq:dT/dt}) one
f\/inds that
\begin{gather*}
{dD\over dt}=0, \qquad \mbox{i.e.} \qquad {d\tau_k\over dt}=0,
\end{gather*}
for all $k=1,\ldots,r$. Obviously all eigenvalues of the
scattering matrix $T(t, \lambda)$ will be expressed in terms of
$\tau_k$. Each $\tau_k(\lambda)$ can be expanded over the negative
powers of $\lambda$:
\[
\tau_k(\lambda)=\sum_{s=0}^{\infty}I_s^{(k)}\lambda^{-s},
\]
and all the coef\/f\/icients $I_s^{(k)}$ will be integrals of
motion.

\section{Real Hamiltonian forms}\label{sec:4}

The Lax representations of the ATFT models (see e.g.
\cite{Mikh,OlPerMikh,Olive,DrSok,SasKha} and the references
therein) are related mostly to the normal real form of the Lie
algebra ${\frak  g} $, see \cite{Helg}.

Our aim here is to:
\begin{enumerate}\vspace{-2mm}
\itemsep=0pt \item[1)] generalize the ATFT to complex-valued
f\/ields $\vec{q\,}^\bbbc = \vec{q\,}^0 + i\vec{q\,}^1$, and to
\item[2)] describe the family of RHF of these ATFT models.
\end{enumerate}

We also provide a tool generalizing of the one in \cite{2} for the
construction of new inequivalent RHF's of the ATFT. The ATFT for
the algebra $sl(n) $ can be written down as an
inf\/inite-dimensional Hamiltonian system as follows:
\begin{gather*}
 {dq_k \over dt } = \{ q_k, H_{\rm ATFT}\}, \qquad {dp_k \over
dt } = \{ p_k,
H_{\rm ATFT}\}, \\
 H_{\rm ATFT} = \int_{-\infty }^{\infty } dx \, \left( {1\over 2 }
(\vec{p}(x,t),\vec{p}(x,t)) + \sum_{k=0}^{r}
e^{-(\vec{q}(x,t),\alpha_k)} \right),
\end{gather*}
where  $\vec{q}(x,t) $ and $\vec{p} = \partial \vec{q}/\partial x
$ are the canonical coordinates  and momenta satisfying canonical
Poisson brackets:
\begin{gather}\label{eq:c-PB}
\{ p_k(x) , q_j(y)\} = \delta_{jk} \delta (x-y).
\end{gather}
Next we def\/ine the involution $\mathcal{ C} $ acting on the
phase space $\mathcal{ M} $ as follows:
\begin{gather*}
 \mbox{1)} \quad \mathcal{ C}(F(p_k,q_k)) = F(\mathcal{ C}(p_k),
\mathcal{ C}(q_k)),  \nonumber\\
 \mbox{2)} \quad \mathcal{ C}\left( \{ F(p_k,q_k), G(p_k,q_k)\}\right) =
\left\{ \mathcal{ C}(F), \mathcal{ C}(G) \right\} , \nonumber\\
 \mbox{3)} \quad \mathcal{ C}(H( p_k,q_k)) = H(p_k,q_k) . \nonumber
\end{gather*}
Here $F(p_k,q_k) $, $G(p_k,q_k) $ and the Hamiltonian $H(p_k,q_k)
$ are functionals on $\mathcal{M} $ depending analytically on the
f\/ields $q_k(x,t) $ and $p_k(x,t) $.

The complexif\/ication of the ATFT is rather straightforward. The
resulting complex ATFT (CATFT) can be written down as standard
Hamiltonian system with twice as many f\/ields $\vec{q\,}^a(x,t)
$, $\vec{p\,}^a(x,t)  $, $a=0,1 $:
\begin{gather*}
\vec{p\,}^\bbbc (x,t) = \vec{p\,}{}^0(x,t)+i \vec{p\,}{}^1(x,t),
\qquad \vec{q\,}^\bbbc (x,t)= \vec{q\,}{}^0(x,t)+i
\vec{q\,}{}^1(x,t),
\\
\big\{{p}_{k}^0(x,t), {q}_{j}^0(y,t)\big \}= -
\big\{{p}_{k}^1(x,t), {q}_{j}^1(y,t) \big\} = \delta _{kj} \delta
(x-y).
\end{gather*}
The densities of the corresponding Hamiltonian and symplectic form
equal
\begin{gather*}
\mathcal{H}_{\rm ATFT}^\bbbc \equiv  \re \mathcal{H}_{\rm ATFT}
(\vec{p\,}{}^0+i \vec{p\,}{}^1, \vec{q\,}{}^0+i \vec{q\,}{}^1) \nonumber\\
\phantom{\mathcal{H}_{\rm ATFT}^\bbbc}{}=  {1\over 2 }
(\vec{p\,}{}^0,\vec{p\,}{}^0) -{1\over 2 }
(\vec{p\,}{}^1,\vec{p\,}{}^1) + \sum_{k=0}^{r}
e^{-(\vec{q\,}{}^0,\alpha _k)}
\cos ((\vec{q\,}{}^1,\alpha _k)) ,  \\
 \omega^\bbbc= (d\vec{p\,}{}^0\wedgecomma
d\vec{q\,}{}^0) - (d\vec{p\,}{}^1\wedgecomma d \vec{q\,}{}^1).
\end{gather*}
The family of RHF then are obtained from the CATFT by imposing an
invariance condition with respect to the involution
$\tilde{\mathcal{ C}} \equiv \mathcal{ C}\circ \ast $ where by
$\ast $ we denote the complex conjugation. The involution
$\tilde{\mathcal{ C}} $ splits the phase space $\mathcal{ M}^\bbbc
$ into a direct sum $\mathcal{ M}^\bbbc \equiv {\cal M}_+^\bbbc
\oplus \mathcal{M}_-^\bbbc$ where
\begin{gather*}
\mathcal{M}_+^\bbbc = \mathcal{ M}_0 \oplus i \mathcal{ M}_1,
\qquad \mathcal{M}_-^\bbbc = i\mathcal{ M}_0 \oplus  \mathcal{
M}_1,
\end{gather*}
The phase space of the RHF is $\mathcal{ M}_\bbbr \equiv
\mathcal{M}_+^\bbbc $.  By $\mathcal{ M}_0 $ and $\mathcal{ M}_1 $
we denote the eigensubspaces of~$\mathcal{ C} $, i.e.\
$\mathcal{C}(u_a)=(-1)^a u_a $ for any $u_a\in {\cal M}_a $.

Then extracting of Real Hamiltonian forms (RHF's) is similar to
the obtaining a real forms of a semi-simple Lie algebra. The
Killing form for the later is indef\/inite in general (it is
negatively-def\/inite  for the compact real forms). So one should
not be surprised of getting RHF's with indef\/inite kinetic energy
quadratic form. Of course this is an obstacle for their
quantization.

Thus to each involution $\mathcal{ C} $ satisfying 1)--3) one can
relate a RHF of the ATFT.  Due to the condition 3) $\mathcal{C} $
must preserve the system of admissible roots of ${\frak g} $; such
involutions can be constructed from the $\bbbz_2 $-symmetries of
the extended Dynkin diagrams of ${\frak  g} $ studied in
\cite{SasKha}.

\section{Examples}\label{sec:5}

In this Section we provide several examples of RHF of ATFT related
to exceptional Kac--Moody algebras with height~1. We show that the
involutions $\mathcal{C} $ in all these cases are dual to $\bbbz_2
$-symmetries  of the extended Dynkin diagrams derived in
\cite{SasKha}.

The examples below illustrate the procedure outlined above and
display new types of ATFT.

\subsection[${\bf E}_6^{(1)}$ Toda field theories]{$\boldsymbol{{\bf E}_6^{(1)}}$
Toda f\/ield theories}\label{ssec:ex1}

\begin{figure}[h!]
\vspace{-4mm} \centerline{
\begin{picture}(400,370)
\put(20,320){$\eddei{\alp{1}}.{\alp{2}}.{\alp{3}}.{\alp{4}}.{\alp{5}}.
{\alp{6}}.{\alp{0}}.\Rightarrow
\eddfid{\bet{0}}.{\bet{2}}.{\bet{4}}.{\bet{3}}.{\bet{1}}.$}
\put(88,281){\vector(-3,1){12}} \put(124,281){\vector(3,1){12}}
\qbezier(88,281)(106,275)(124,281) \put(55,281){\vector(-3,1){12}}
\put(157,281){\vector(3,1){12}} \qbezier(55,281)(106,264)(157,281)
\end{picture}}
\vspace{-3.8in} \caption{${\bf E}_{6}^{(1)}\rightarrow {\bf
F}_4^{(1)}$.} \label{fig:1}
\end{figure}
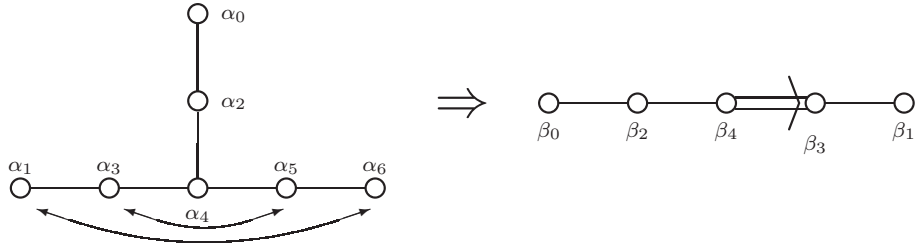

The set of admissible roots for this algebra is
\begin{gather*}
\alpha_1={1\over 2}(e_1-e_2-e_3-e_4-e_5-e_6-e_7+e_8), \qquad
\alpha_2=e_1+e_2, \\
\alpha_3=e_2-e_1,\qquad \alpha_4=e_3-e_2, \qquad
\alpha_5=e_4-e_3, \qquad \alpha_6=e_5-e_4,\nonumber\\
\alpha_0=-{1\over 2}(e_1+e_2+e_3+e_4+e_5-e_6-e_7+e_8),
\end{gather*}
where $\alpha_1,\dots , \alpha_6$ form the set of simple roots of
${\bf E}_6$ and $\alpha_0$ is the minimal root of the algebra.
This is the standard def\/inition of the root system of ${\bf E}_6
$ embedded into the 8-dimensional Euclidean space $\bbbe^8 $. The
root space $\bbbe_6 $ of the algebra ${\bf E}_6$ is the
6-dimensional subspace of $\bbbe^8 $ orthogonal to the vectors
$e_7+e_8 $ and $e_6+e_7+2e_8 $. Thus any vector $\vec{q} $
belonging to $\bbbe_6 $ has only 6 independent coordinates and can
be written as:
\begin{gather}\label{eq:vec-q}
\vec{q} = \sum_{k=1}^{5} q_k e_k + q_6 e'_6, \qquad e'_6 = {1\over
\sqrt{3} } (e_6 + e_7 - e_8).
\end{gather}
Let us f\/ix up the action of the involution $\mathcal{C} $ on a
generic vector $\vec{q} $ in $\bbbe^8 $ by:
\begin{gather}
\mathcal{C} (q_k) = -q_{5-k}+{1\over 2}\sum_{m=1}^4 q_{m}, \qquad
\mbox{for \; } k=1,\dots ,4 , \nonumber\\
\phantom{\mathcal{C} (q_k)}{}= q_{13-k}-{1\over 2}\sum_{m=5}^8
q_{m},\qquad \mbox{for \; } k=5,\dots ,8.\label{eq:e_6}
\end{gather}
This action is compatible with the $\bbbz_2 $-symmetry
$\mathcal{C}^\# $ of the extended Dynkin diagram (see
Fig.~\ref{fig:1})  and ref\/lects an involution of the Kac--Moody
algebra ${\bf E}_{6}^{(1)} $, see~\cite{Kac}.
 It acts on the root space as follows:
\begin{gather}
\mathcal{C}^\# e_k = -e_{5-k}+{1\over 2}\sum_{m=1}^4e_{m}, \qquad
\mbox{for \;} k=1,\dots ,4, \nonumber\\
\phantom{\mathcal{C}^\# e_k}{}= e_{13-k}-{1\over
2}\sum_{m=5}^8e_{m}, \qquad
\mbox{for \;} k=5,\dots ,8 , \nonumber\\
\mathcal{C}^\#\alpha _1 =\alpha _{6}, \qquad \mathcal{C}^\#\alpha
_3 =\alpha _{5}, \qquad \mathcal{C}^\#\alpha _k =\alpha _{k},
\qquad k=0,2,4. \label{eq:E1.2}
\end{gather}

The involution $\mathcal{C}^\# $ splits the root space $\bbbe_{6}
$ into a direct sum of its eigensubspaces:
$\bbbe_{6}=\bbbe_+\oplus \bbbe_- $ with $\dim \bbbe_+=4$, $\dim
\bbbe_-=2 $. The vectors:
\begin{alignat*}{4}
&\widetilde{e}_1 = {1\over 2} (e_5 - \sqrt{3} e'_6), \quad &&
\widetilde{e}_2 = {1\over 2} (e_1+e_2+e_3+e_4),& & \widetilde{e}_3 = {1\over 2} (-e_1-e_2+e_3+e_4),& \nonumber\\
& \widetilde{e}_4 = {1\over 2} (-e_1+e_2-e_3+e_4),\quad & 
& \widetilde{e}_5 = {1\over 2} (-e_1+e_2+e_3-e_4), \quad &&
\widetilde{e}_6 = {1\over 2} (\sqrt{3} e_5 + e'_6). & 
\end{alignat*}
form an orthonormal basis in $\bbbe_6 $. The f\/irst four satisfy
$\mathcal{C}^\# \widetilde{e}_k =\widetilde{e}_k$, $k=1,\dots, 4
$, so they span~$\bbbe_+ $; the last two  span $\bbbe_- $ because
$\mathcal{C}^\# \widetilde{e}_j =-\widetilde{e}_j$, $j=5,6$. In
terms of $\widetilde{e}_k $ the admissible root system of ${\bf
F}_4^{(1)} $ takes the standard form:
\begin{gather*}
\beta _0 = -\widetilde{e}_2 - \widetilde{e}_1, \qquad \beta _1 =
{1\over 2}(\widetilde{e}_1 -\widetilde{e}_2 - \widetilde{e}_3
-\widetilde{e}_4), \nonumber\\
\beta _2 = \widetilde{e}_2 - \widetilde{e}_3,\qquad \beta _3 =
\widetilde{e}_4, \qquad \beta_4 =  \widetilde{e}_3 -
\widetilde{e}_4.
\end{gather*}
satisfying $\beta _0+2\beta _1+2\beta _2+4\beta _3+3\beta _4=0 $.

Let us take the complex vector $\vec{q}(x,t) = \vec{q\,}^0(x,t) +
i\vec{q\,}^1(x,t)\in \bbbe_6 $ (i.e., of the form
(\ref{eq:vec-q})) and let $\vec{p}(x,t) = \partial
\vec{q}/\partial x $. Let us denote their projections onto
$\bbbe_\pm $ by $\vec{q}_\pm $ and $\vec{p}_\pm $ respectively.
Then the densities $\mathcal{H}_{1}^\bbbr$, $\omega _{1}^\bbbr$
for the RHF of AFTF equal:
\begin{gather*}
\mathcal{H}_{1}^\bbbr  = {1\over 2} \left(
(\vec{p\,}^0_+(x,t),\vec{p\,}^0_+(x,t)) -
(\vec{p\,}^0_-(x,t),\vec{p\,}^0_-(x,t))\right)
+  {\rm e}^{-(\vec{q\,}^0_+(x,t),\beta _0)} \nonumber\\
\phantom{\mathcal{H}_{1}^\bbbr  =}{} +2 {\rm
e}^{-(\vec{q\,}^0_+(x,t),\beta _1)}
\cos((\vec{q\,}^1_-(x,t),\widetilde{e}_5 + \sqrt{3}\widetilde{e}_6
)) + 2 {\rm e}^{-(\vec{q\,}^0_+(x,t),\beta _2)}  \\
\phantom{\mathcal{H}_{1}^\bbbr  =}{}+  4 {\rm
e}^{-(\vec{q\,}^0_+(x,t),\beta _3)}
\cos((\vec{q\,}^1_-(x,t),\widetilde{e}_5) ) + 3 {\rm e}^{-(\vec{q\,}^0_+(x,t),\beta _4)} , \nonumber\\
 \omega _{1}^\bbbr = \big( \delta
\vec{p}_+(x) \wedgecomma \delta \vec{q}_+(x)\big) -\big( \delta
\vec{p}_-(x) \wedgecomma \delta \vec{q}_-(x)\big) ,
\end{gather*}
If we put $\vec{q}_-(x,t) = 0 $ then also $\vec{p}_-(x,t) = 0 $
and we get the reduced ATFT related to the Kac--Moody algebra
${\bf F}_{4}^{(1)} $ \cite{SasKha}.

\subsection[${\bf E}_7^{(1)}$ Toda f\/ield theories]{$\boldsymbol{{\bf E}_7^{(1)}}$
 Toda f\/ield theories}\label{ssec:ex2}

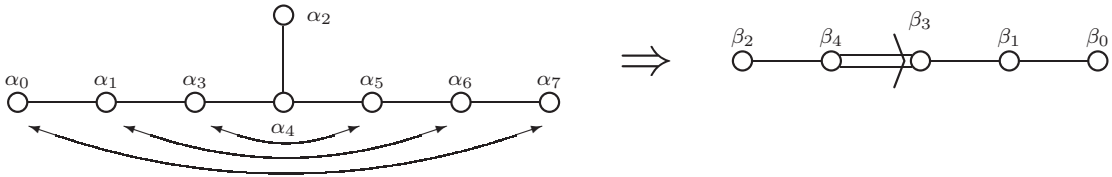
\begin{figure}[h!]
\vspace{-5mm} \centerline{\hspace{-25mm}\begin{picture}(400,520)
\put(5,540){}
\put(20,480){$\eddeii{\alp{1}}.{\alp{2}}.{\alp{3}}.{\alp{4}}.{\alp{5}}.
{\alp{6}}.{\alp{7}}.{\alp{0}}.~\Rightarrow~
\eddfiiu{\bet{2}}.{\bet{4}}.{\bet{3}}.{\bet{1}}.{\bet{0}}.$}
\put(225,457){\vector(3,1){10}} \put(51,457){\vector(-3,1){10}}
\qbezier(51,457)(138,428)(225,457) \put(189,457){\vector(3,1){10}}
\put(87,457){\vector(-3,1){10}} \qbezier(189,457)(138,440)(87,457)
\put(156,457){\vector(3,1){10}} \put(120,457){\vector(-3,1){10}}
\qbezier(120,457)(138,451)(156,457)
\end{picture}}
\vspace{-6.1in} \caption{The ${\bf E}_{7}^{(1)}~\rightarrow~ {\bf
E}_{6}^{(2)}$ RHF of af\/f\/ine TFT. \label{fig:E7}} \end{figure}

The set of admissible roots for this algebra is
\begin{gather*}
\alpha_1={1\over 2}(e_1-e_2-e_3-e_4-e_5-e_6-e_7+e_8), \qquad
\alpha_2=e_1+e_2, \\ \alpha_3=e_2-e_1,\qquad \alpha_4=e_3-e_2,
\qquad
\alpha_5=e_4-e_3, \qquad \alpha_6=e_5-e_4,\nonumber\\
\alpha_7=e_6-e_5, \qquad \alpha_0=e_7-e_8,
\end{gather*}
where $\alpha_1,\dots , \alpha_7$ form the set of simple roots of
${\bf E}_7$ and $\alpha_0$ is the minimal root of the algebra.
This is the standard def\/inition of the root system of ${\bf E}_7
$ embedded into the 8-dimensional Euclidean space $\bbbe^8 $. The
root space $\bbbe_7 $ of the algebra ${\bf E}_7$ is the
7-dimensional subspace of $\bbbe^8 $ orthogonal to the vector
$e_7+e_8 $. Thus any vector $\vec{q} $ belonging to $\bbbe_7 $ has
7 independent coordinates and can be written as:
\begin{gather}\label{eq:vec-q'}
\vec{q} = \sum_{k=1}^{6} q_k e_k + q_7 e'_7, \qquad e'_7 = {1\over
\sqrt{2} } (e_7 - e_8).
\end{gather}
Let us f\/ix up the action of the involution $\mathcal{C} $ on a
generic vector $\vec{q} $ in $\bbbe^8 $ by equation
(\ref{eq:e_6}). This action is compatible with the $\bbbz_2
$-symmetry $\mathcal{C}^\# $ of the extended Dynkin diagram (see
Fig.~\ref{fig:E7})  and ref\/lects an involution of the Kac--Moody
algebra ${\bf E}_{7}^{(1)} $, see~\cite{Kac}.  It acts on the root
space as in equation (\ref{eq:E1.2}) above.

However its action on the root space $\bbbe_{7} $ is dif\/ferent.
It splits $\bbbe_7 $ into a direct sum of its eigensubspaces:
$\bbbe_{7}=\bbbe_+\oplus \bbbe_- $ with $\dim \bbbe_+=4$, $\dim
\bbbe_-=3 $. The vectors:
\begin{gather*}
\widetilde{e}_1 = {1\over 2\sqrt{2}} (e_1+e_2+e_3+e_4+e_5 - e_6
-\sqrt{2} e'_7), \qquad
\widetilde{e}_3 = {1\over \sqrt{2}} (-e_1+e_4), \nonumber\\
\widetilde{e}_2 = {1\over 2\sqrt{2}} (-e_1-e_2-e_3-e_4+e_5 - e_6
-\sqrt{2} e'_7), \qquad \widetilde{e}_4 = {1\over \sqrt{2}}
(-e_2+e_3),
\end{gather*}
form an orthonormal basis in $\bbbe_+ $. Indeed, they satisfy
$(\widetilde{e}_j, \widetilde{e}_k)=\delta _{jk} $ and
$\mathcal{C}^\# \widetilde{e}_k =\widetilde{e}_k$, $k=1,\dots, 4
$. In terms of $\widetilde{e}_k $ the admissible root system of
${\bf E}_6^{(2)} $ takes the standard form:
\begin{gather*}
\beta _0 = -\widetilde{e}_2 - \widetilde{e}_1, \qquad \beta _1 =
\widetilde{e}_2 - \widetilde{e}_3, \qquad
\beta _2 = \widetilde{e}_3 - \widetilde{e}_4, \nonumber\\
\beta _3 = 2\widetilde{e}_4, \qquad  \beta _4 = \widetilde{e}_1
-\widetilde{e}_2 - \widetilde{e}_3 -\widetilde{e}_4,
\end{gather*}
satisfying $\beta _0+2\beta _1+\beta _2+3\beta _3+2\beta _4=0 $.

The vectors
\begin{gather*}
\widetilde{e}_5 = {1\over 2} (-e_1+e_2+e_3-e_4), \qquad
\widetilde{e}_6 = {1\over \sqrt{2}} (e_5 + e_6),\qquad
\widetilde{e}_7 = {1\over 2} (-e_5+e_6-\sqrt{2}e'_7),
\end{gather*}
span $\bbbe_- $ because $\mathcal{C}^\# \widetilde{e}_j
=-\widetilde{e}_j$, $j=5,6,7$.

Let us take the complex vector $\vec{q}(x,t) = \vec{q\,}^0(x,t) +
i\vec{q\,}^1(x,t)\in \bbbe_7 $ (i.e., of the form
(\ref{eq:vec-q'})) and let $\vec{p}(x,t) = \partial
\vec{q}/\partial x $. Let us denote their projections onto
$\bbbe_\pm $ by $\vec{q}_\pm $ and $\vec{p}_\pm $ respectively.
Then the densities $\mathcal{H}_{1}^\bbbr$, $\omega _{1}^\bbbr$
for the RHF of AFTF equal:
\begin{gather*}
\mathcal{H}_{2}^\bbbr  = {1\over 2} \left(
(\vec{p\,}^0_+(x,t),\vec{p\,}^0_+(x,t)) -
(\vec{p\,}^0_-(x,t),\vec{p\,}^0_-(x,t))\right) +  2{\rm
e}^{-(\vec{q\,}^0_+(x,t),\beta _0)}
\cos((\vec{q\,}^1_-(x,t),\widetilde{e}_7)) \nonumber\\
\phantom{\mathcal{H}_{2}^\bbbr  =}{}+4 {\rm
e}^{-(\vec{q\,}^0_+(x,t),\beta _1)}
\cos\left((\vec{q\,}^1_-(x,t),\frac{1}{2}(\widetilde{e}_5+
\sqrt{2}\widetilde{e}_6-\widetilde{e}_7)\right)+ 2 {\rm
e}^{-(\vec{q\,}^0_+(x,t),
\beta _2)}  \\
\phantom{\mathcal{H}_{2}^\bbbr  =}{}+  6 {\rm
e}^{-(\vec{q\,}^0_+(x,t),\beta _3)}
\cos((\vec{q\,}^1_-(x,t),\widetilde{e}_5) ) + 4 {\rm
e}^{-(\vec{q\,}^0_+(x,t), \beta _4)} , \nonumber\\
\omega _{2}^\bbbr = \big( \delta \vec{p}_+(x)\wedgecomma \delta
\vec{q}_+(x)\big) -\big( \delta \vec{p}_-(x)\wedgecomma \delta
\vec{q}_-(x)\big) .
\end{gather*}
Again, if we put $\vec{q}_-(x,t) = 0 $ then also $\vec{p}_-(x,t) =
0 $ and we get the reduced ATFT related to the Kac--Moody algebra
${\bf E}_{6}^{(2)} $~\cite{SasKha}.

\section[Classical $R $-matrix method and ATFT]{Classical $\boldsymbol{R}$-matrix method and ATFT}\label{sec:5a}

There are several methods to approach the Hamiltonian properties
of the ATFT, see e.g.\ \cite{Bogo,DrSok,Como,Dam1,Dam2}. One of the
ef\/fective methods is based on the well known classical $R
$-matrix \cite{PPK} which is introduced by:
\begin{gather}\label{eq:R-mat}
\big\{ U(x,\lambda )\otimescomma U(y,\mu )\big\} = \left[
R(\lambda ,\mu ) , U(x,\lambda )\otimes \openone +\openone \otimes
U(y,\mu ) \right] \delta (x-y),
\end{gather}
where $\big\{ U(x,\lambda )\otimescomma U(y,\mu )\big\}_{ij,kl} =
\{ U_{ij}(x,\lambda ), U_{kl}(y,\mu )\}. $ In order to apply this
def\/inition ef\/fecti\-ve\-ly we make use of another Lax operator for
the ATFT:
\begin{gather*}
\widetilde{\widetilde{L}} \widetilde{\widetilde{\psi }}(x,\lambda
) \equiv  i {d \widetilde{\widetilde{\psi }} \over dx  } +
U(x,\lambda )
\widetilde{\widetilde{\psi }}(x,\lambda ) =0, \\
U(x,\lambda ) = -{i \over 2} \sum_{j=1}^{r}q_{j,x} H_j - \lambda
\sum_{j=0}^{r} e^{-(\alpha _j,\vec{q})/2} E_{\alpha _j}, \nonumber
\end{gather*}
which is gauge equivalent to $L $ (\ref{eq:2.1}). Since the
Poisson brackets are introduced by (\ref{eq:c-PB}) the left hand
side of (\ref{eq:R-mat}) takes the form:
\begin{gather*}
\big\{U_{ij}(x,\lambda )\otimescomma U_{kl}(y,\mu )\big\} = {i
\over 4} \sum_{k=0}^{r} e^{-(\alpha _k,\vec{q})/2} \left( \mu
H_{\alpha _k} \otimes E_{\alpha _k} - \lambda E_{\alpha _k}
\otimes H_{\alpha _k} \right)\delta(x-y).
\end{gather*}
Thus equation (\ref{eq:R-mat}) becomes an over-determined set of
equations for $ R(\lambda ,\mu ) $ which is solved by~\cite{PPK}:
\begin{gather*}
R(\lambda ,\mu ) = {1 \over 4i} {\lambda ^h + \mu ^h\over
\lambda ^h - \mu ^h } \sum_{k=1}^{r} H_k\otimes H_k +{1 \over 2i}
\sum_{\alpha \in \Delta } \left( {\lambda \over \mu
}\right)^{p(\alpha )} {\mu ^h E_\alpha \otimes
E_{-\alpha } \over \lambda ^h - \mu ^h} \nonumber\\
\phantom{R(\lambda ,\mu )}{}= {1 \over 4i\sinh (h\eta/2) } \left( \cosh (h\eta/2)
\sum_{k=1}^{r} H_k\otimes H_k +\sum_{\alpha \in \Delta }
e^{(p(\alpha )-h/2)\eta} E_\alpha \otimes E_{-\alpha } \right),
\end{gather*}
where $h $ is the Coxeter number of $\mathfrak{g} $, $\eta =\ln
(\lambda /\mu ) $ and $p(\alpha ) $ is the height of the root
$\alpha  $ modulo $h $.  If $\alpha =\sum\limits_{j=1}^{r}
n_{\alpha ,j} \alpha _j $ where $n_{\alpha ,j} $ are integers,
then $p(\alpha ) =\sum\limits_{j=1}^{r}n_{ \alpha ,j} $.

The relation (\ref{eq:R-mat}) allows one to derive the Poisson
brackets between the matrix elements of the fundamental solutions
$T^+_{x_0}(x,\lambda ) $ and $T^-_{x_0}(x,\lambda ) $ (or the
scattering matrix $T_{x_0}(\lambda ) $) of $L $ which are
def\/ined by:
\begin{gather*}
LT_{x_0}^\pm (x,\lambda )= 0 , \qquad \lim_{x\to \pm x_0}
T^\pm _{x_0}(x,\lambda ) e^{-i\lambda xJ_0} =\openone , \\
T_{x_0}(\lambda ) = (T^+_{x_0}(x,\lambda ))^{-1}
T^-_{x_0}(x,\lambda ).
\end{gather*}
Then
\begin{gather*}
\big\{ T^\pm _{x_0}(x,\lambda ) \otimescomma T^\pm _{x_0}(y,\mu )
\big\} = \left[ R(\lambda ,\mu ), T^\pm_{x_0}(x,\lambda )\otimes
T^\pm_{x_0}(y,\mu )\right],
\end{gather*}
and
\begin{gather}\label{eq:Pb-T-l}
\big\{ T_{x_0}(\lambda ) \otimescomma T_{x_0} (\mu ) \big\} =
\left[ R(\lambda ,\mu ), T_{x_0}(\lambda )\otimes  T_{x_0}(\mu
)\right],
\end{gather}
These results hold true for potentials on compact support provided
we choose $x_0 $ large enough so that $\vec{q}_x=0 $ for $|x|>x_0
$.

With $T_{x_0}(t,\lambda)$ we can associate a set of generating
functionals $\tau_k(x_0,\lambda)$ of the integrals of motion
\begin{gather*}
T_{x_0}(t, \lambda
)=u_0^{-1}(x_0,t,\lambda)D_{x_0}(\lambda)u_0(x_0,t,\lambda),
\qquad D_{x_0}(\lambda)=\exp \left(\sum_{k=1}^{r}
{2\tau_k(x_0,\lambda)\over (\alpha_k,\alpha_k)}H_{\alpha_k}
\right).
\end{gather*}
Again we can write the expansion
\[
\tau_k(x_0,\lambda)=\sum_{s=0}^{\infty}I_{x_0,k}^{(s)}\lambda^{-s}.
\]
An important consequence of (\ref{eq:Pb-T-l}) is that the
functions $\tau_k(x_0,\lambda)$ are in involution. Indeed
from~(\ref{eq:Pb-T-l}) there follows that:
\begin{gather}\label{eq:tr-Tk}
\{ \mbox{tr}\, T^k_{x_0}(\lambda), \mbox{tr}\, T^p_{x_0}(\mu)\}=0,
\end{gather}
for any pair of integers $k, p$. Obviously $\mbox{tr}\,
T^k(\lambda)$ can be expressed in terms of the
invariants~$\tau_k(\lambda)$~(\ref{eq:D-tau}) of the scattering
matrix $T(\lambda) \in \mathcal{G}$. Therefore from
(\ref{eq:tr-Tk}) we f\/ind that
\begin{gather*}
\{ \tau_k(x_0,\lambda), \tau_m(x_0,\mu)\}=0,\qquad 1\leq k,m \leq
r,
\end{gather*}
i.e.\ the integrals of motion $I^{(s)}_k$ are all in involution:
\begin{gather*}
\{ I^{(s)}_{x_0,k}, I^{(n)}_{x_0,m}\}=0,\qquad  0 \leq k,m \leq r,
\quad s,n \geq 0.
\end{gather*}
In order to derive the corresponding results for the RHF of the
ATFT we have to use the fact that the automorphism $\mathcal{C} $
induces an automorphism $\mathcal{C}^\vee $ on the Lie algebra
$\mathfrak{g} $ and on the Lax operator as follows \cite{GeVi}:
\begin{gather*}
\mathcal{C}^\vee (H_j) = H_{\mathcal{C}^\#(e_j)}, \qquad
\mathcal{C}^\vee E_{\alpha } = E_{\mathcal{C}^\#(\alpha )},\qquad
L(\mathcal{C}(\vec{q}^*),x,\lambda^* ) = \mathcal{C}^\vee (
L(\vec{q},x,\lambda ))^*.
\end{gather*}
These relations allow one to prove that the fundamental solution
$T_{x_0} (x,\lambda ) $ and the scattering matrix $T_{x_0}(\lambda
) $ for the corresponding RHF model has the properties:
\begin{gather*}
(T_{x_0}(x,\lambda^* ) )^* = \mathcal{C}^\vee (T_{x_0}(x,\lambda
)), \qquad (T_{x_0}(\lambda^* ) )^* = \mathcal{C}^\vee
(T_{x_0}(\lambda )),
\end{gather*}
and as a consequence: $ (\tau_k(\lambda ^*)) ^* =
\tau_{\bar{k}}(\lambda))$, where $\bar{k} $ is def\/ined through
$\mathcal{C}^\# (\alpha _k) = \alpha _{\bar{k}}$.

\section{Conclusions}\label{sec:6}

The RHF of the ATFT  models related to the exceptional Kac--Moody
algebras ${\bf E}_6^{(1)}$  and ${\bf E}_7^{(1)}$ are constructed.
These  models generalize the ones in \cite{SasKha} since they
contain two types of f\/ields $\vec{q}_+(x,t) $ and
$\vec{q}_-(x,t) $ with dif\/ferent properties with respect to the
involution $\mathcal{C} $. The models in~\cite{SasKha} contain
only f\/ields invariant with respect to $\mathcal{C} $.

We outlined the derivation of the Hamiltonian properties through
the classical $R$-matrix approach.

\subsection*{Acknowledgments}

One of us (GGG) thanks the organizing committee of the Sixth
International Conference ``Symmetry in Nonlinear Mathematical
Physics''  for the scholarship and for the warm hospitality in
Kyiv. The present paper is the written version of the talk
delivered by GGG at this conference. The work of GGG is supported
by the Bulgarian National Scientif\/ic Foundation Young Scientists
Scholarship for the project ``Solitons, Dif\/ferential Geometry
and Biophysical Models''. We also acknowledge support by the
National Science Foundation of Bulgaria, contract No. F-1410. We
also thank an anonymous referee for careful reading of the
manuscript and for useful suggestions.

\LastPageEnding

\end{document}